\documentclass[aps,pre,reprint,10pt,superscriptaddress,showpacs]{revtex4-1}
\usepackage{amsfonts}
\usepackage{amsmath}
\usepackage{amssymb}
\usepackage{graphicx}
\usepackage{subfigure}
\usepackage{color}
\begin{document}
\title{Landau damping effects on dust-acoustic solitary waves in a dusty negative-ion plasma}
\author{Arnab Barman}
\author{A. P. Misra}
\email{apmisra@visva-bharati.ac.in; apmisra@gmail.com}
\affiliation{Department of Mathematics, Siksha Bhavana, Visva-Bharati University, Santiniketan-731 235, West Bengal, India}
\pacs{52.25 Dg, 52.27.Cm, 52.35.Mw, 52.35.Sb}
\begin{abstract}
The nonlinear theory of dust-acoustic waves (DAWs) with Landau damping is studied in an unmagnetized  dusty negative-ion plasma in the extreme conditions when the free electrons are absent. The cold massive charged dusts are described by fluid equations, whereas the two-species of ions (positive and negative) are described by the kinetic Vlasov equations. A Korteweg de-Vries (KdV) equation with Landau damping,  governing the dynamics of  weakly nonlinear and weakly dispersive DAWs,  is derived following   Ott and Sudan [Phys. Fluids {\bf 12}, 2388 (1969)]. It is shown that for some typical laboratory and space plasmas, the Landau damping (and the nonlinear) effects are more pronounced than the finite Debye length (dispersive) effects for which the KdV soliton theory is not applicable to DAWs in dusty pair-ion plasmas.    The properties of the linear phase velocity, solitary wave amplitudes (in presence and absence of the Landau damping) as well as the Landau damping rate are studied with the effects of the positive ion to dust density ratio $(\mu_{pd})$ as well as the ratios of  positive   to negative ion temperatures $(\sigma)$ and masses $(m)$.
\end{abstract}
\maketitle
\section{Introduction}
Plasmas with massive charged dust grains are of interest both for space (e.g., cometary tails, planetary rings, the interstellar medium, Earth's magnetosphere and upper atmosphere, Earth's D and lower E regions etc.) \cite{shukla2002,shukla2002} as well as laboratory plasmas \cite{geortz1989,merlino1998}. Furthermore, dusty plasmas containing both positive and negative ions and a very few percentage of free electrons (or almost electron-free) are found in both naturally occurring plasmas (e.g., Earth's D and lower E regions, the F-ring of Saturn, nighttime polar mesosphere etc.) \cite{geortz1989,narcisi1971,rapp2005} and in plasmas used for technological applications \cite{choi1993,yabe1994}. There are several motivations for investigating such plasmas and associated waves and instabilities owing to their potential applications in both laboratory and space plasma environments (See, e.g., Refs. \onlinecite{kim2006,kim2013,rosenberg2007,misra2012,misra2013,rehman2012,ghosh2013,oohara2005,saleem2007}). In dusty plasmas, the size of the dust grains may vary in the range of $0.05-10\mu$m, their mass is about $10^6-10^{12}$ times the mass of ions and they have atomic numbers $z_d$ in the range of $10^3-10^5$ \cite{shukla2002}.  In typical dusty plasmas with electrons and ions,  dust grains are  negatively charged due to high mobility of electrons into dust grain surface \cite{shukla2002}. However, recent laboratory experiments \cite{kim2006,kim2013,rosenberg2007} suggest that in dusty negative-ion plasmas where positive ions are the more mobile species, dusts can be positively charged when  $m_n>m_p$, $n_n\gg n_e$ and $T_p>T_n$  are satisfied, where $m_j$ is the mass, $n_j$ is the number density and $T_j$ is the thermodynamic temperature of $j$-species particles in which $j=e,p,n$ stand for electrons, positive ions and negative ions respectively. On the other hand, when charged dust grains collect all the electrons from the background plasma, they can be negatively charged. Such density depletion of electrons associated with the capture of electrons by aerosol particles have been observed in the summer polar mesosphere at about 85 km altitude \cite{reid1990}. Furthermore,  positively charged nanometer-seized particles were observed in the nighttime polar mesosphere in a region (Altitude range between 80 and 90 km) dominated by positive and negative ions and a very few percentage of electrons \cite{rapp2005}. 
 
It has been found that the presence of charged dust grains modifies the plasma wave phenomena and these charged dust grains give  rise to new low-frequency eigenmodes, called dust-acoustic wave (DAW), which was first theoretically predicted by 
Rao \textit{et al.} \cite{rao1990}.   However, the presence of negative ions in dusty plasmas can significantly modify not only the dispersion properties of DAWs but also some nonlinear localized structures (For some recent theoretical developments and experiments in negative ion plasmas readers are referred to Refs. \onlinecite{misra2012,misra2013,rehman2012,ghosh2013}). On the other hand, these waves can be damped (collisionless) due to the resonance of particles (trapped and/or free) with the wave (i.e., when the particle's velocity is nearly equal to the wave phase velocity) \cite{landau1946}. The linear electron Landau damping of nonlinear ion-acoustic solitary waves was first  studied by Ott and Sudan \cite{ott1969,ott1970} neglecting the particle's trapping effects  on the assumption that the particle trapping time is much longer than that of Landau damping. They derived a Korteweg-de Vries (KdV) equation with  a source term that models the lowest-order effects of resonant particles. It was demonstrated that an initial wave form may either steepen or not depending on the relative size of the nonlinearity compared to the Landau damping. The latter  was also shown to cause decay of wave amplitude with time.

In the past,  several authors have attempted to study the effects of Landau damping on the nonlinear propagation of electrostatic solitary waves in different plasma systems. For example, Bandyopadhyay {\it et al.} \cite{bandyo2002a,bandyo2002b} investigated the propagation characteristics of ion-acoustic solitary waves with Landau damping in nonthermal plasmas. Ghosh {\it et al.} \cite{ghosh2011} considered the linear Landau damping effects on nonlinear ion-acoustic solitary waves in electron-positron-ion plasmas.  They studied the properties of the wave phase velocity, the solitary wave amplitude as well as the Landau damping rate with the important effects of positron density and temperature. Motivated by the recent theoretical developments as well as experimental observations  of low-frequency electrostatic   waves in pair-ion plasmas \cite{kim2006,kim2013,rosenberg2007,misra2012,misra2013,rehman2012,ghosh2013,oohara2005,saleem2007}  we  have investigated the Landau damping effects of dust-acoustic (DA) solitary waves in dusty pair-ion plasmas (quite distinctive from electron-positron-ion plasmas and dusty electron-ion plasmas).  We  show that for typical plasma parameters relevant for laboratory and space environments, the Landau damping (and the nonlinear) effect is stronger than the finite Debye length (dispersive) effects for which the KdV soliton theory is not applicable to DAWs in dusty pair-ion plasmas.  The landau damping effect is shown to slow down the wave amplitude with time.  It is found that in contrast to electron-ion or electron-positron-ion plasmas, the nonlinearity in the KdV equation can never vanish in dusty pair-ion plasmas with positively or negatively charged dusts. This implies that the modified KdV equation is no longer required for the evolution of DAWs.  The properties of the linear phase velocity, solitary wave amplitudes (in presence and absence of the Landau damping) as well as the Landau damping rate are also analyzed with the effects of positive ion to dust density ratio $(\mu_{pd})$ as well as the ratios of   positive   to negative ion temperatures $(\sigma)$ and masses $(m)$.
\section{Basic Equations}      
    We consider the nonlinear propagation of electrostatic DAWs in an unmagnetized collisionless dusty plasma consisting of singly charged adiabatic positive and negative ions, and positively or negatively charged mobile dusts. The latter are assumed to have equal mass and charge which are treated as constant. It is to be noted that the dust charge fluctuation process may introduce a new low-frequency wave eigen mode as well as  another dissipative effect (wave damping) into the system. However, we have neglected this charge fluctuation effect on the propagation of dust-acoustic waves by the assumption that the charging rate of dust grains is very high compared to the dust plasma oscillation frequency.    The collisions of all particles are also neglected in the considered interval of time. Furthermore, in dusty plasmas the ratio of electric charge to mass of dust grains remains much smaller than those of both positive and negative ions. It is also assumed that the size of the grains is small compared to the average  distance between them.  The unperturbed state is  overall neutral so that the internal electric field is zero. Also,  the curl of the electric field vanishes (i.e., electrostatic), and the perturbation about the equilibrium state is weak.   At equilibrium, the overall charge neutrality condition reads 
    \begin{equation}
    n_{p0}+\zeta z_dn_{d0}=n_{n0}, \label{charge-neutrality}
    \end{equation}
 where $n_{j0}$ is the unperturbed number density of species $j$ ($j$=$p$, $n$, $d$ respectively stand for positive ions, negative ions, and dynamical charged dusts), $z_d$ ($>0$) is the unperturbed  dust charge state and $\zeta=\pm1$ according to when    dusts are positively or negatively charged.  
 
    The cold dust   is described by a set of fluid equations \eqref{cont-eqn}-\eqref{montm-eqn}, whereas the two species of ions (positive and negative) are described by the kinetic Vlasov equation \eqref{Vlasov-eqn}. The electric potential $\phi$ is described by the Poisson equation \eqref{Poisson-eqn}. Thus, the basic equations for the dynamics of charged particles in one space dimension are 
   \begin{equation}
   \frac{\partial n_d}{\partial t}+\frac{\partial (n_du_d)}{\partial x}=0, \label{cont-eqn}
   \end{equation}
   \begin{equation}
   \frac{\partial u_d}{\partial t}+u_d\frac{\partial u_d}{\partial x}=-\frac{q_d}{m_d}\frac{\partial \phi}{\partial x}, \label{montm-eqn}
   \end{equation}
   \begin{equation}
   \frac{\partial f_j}{\partial t}+v\frac{\partial f_j}{\partial x}-\frac{q_j}{m_j}\frac{\partial \phi}{\partial x}\frac{\partial f_j}{\partial v}=0, \label{Vlasov-eqn}
   \end{equation}
   \begin{equation}
   \frac{\partial^2\phi}{\partial x^2}=-4\pi e(n_p-n_n+\zeta z_dn_d), \label{Poisson-eqn}
   \end{equation}
 where $j$ stands for $p$, $n$ denoting, respectively, the positive and negative ions. The ion densities are given by 
 \begin{equation}
 n_j=\int_{-\infty}^{\infty} f_j dv. \label{density-eqn}
 \end{equation}
 In equations \eqref{cont-eqn}-\eqref{density-eqn} $n_d$, $u_d$, $q_d(=\pm z_de)$, $m_d$, respectively, denote the number density, fluid velocity, charge and mass of dust grains. Also, $v$ is the particle's velocity, and $f_j$, $m_j$ and $n_j$, respectively, denote the   velocity distribution function, mass and number densities of $j$-species ions. Furthermore,  $\phi$ is the electrostatic potential and $x$ and $t$ are the space and time coordinates. 

   Equations \eqref{cont-eqn}-\eqref{density-eqn} can be recast in terms of dimensionless variables. We normalize the physical quantities according to $u_d\rightarrow u_d/c_s$, $\phi\rightarrow e\phi/k_BT_p$, $n_j\rightarrow n_j/n_{j0}$, $n_d\rightarrow n_d/n_{d0}$, $f_j\rightarrow f_jv_{tj}/n_{j0}$, $v\rightarrow v/v_{tp}$, where $c_s=\sqrt {z_dk_BT_p/m_d}=\omega_{pd}\lambda_D$ is the DA speed with $\omega_{pd}=\sqrt{4\pi n_{d0}z^2_d e^2/m_d}$ and $\lambda_D=\sqrt{k_BT_p/4\pi n_{d0}z_d e^2}$ denoting, respectively, the dust plasma frequency and the plasma Debye length. Here, $k_B$ is the Boltzmann constant, $T_j$ is the thermodynamic temperature   and $v_{tj}(=\sqrt{k_BT_j/m_j})$ is the thermal velocity of  $j$-species ions. The space and time variables are normalized by $L$ and $L/c_s$ respectively, where $L$ is the characteristic scale length for variations of $n_j,~u_d,~\phi,~f_j$ etc.   
   
   Thus, from equations \eqref{cont-eqn}-\eqref{density-eqn}, we obtain the following set of equations in dimensionless form: 
   \begin{equation}
   \frac{\partial n_d}{\partial t}+\frac{\partial (n_du_d)}{\partial x}=0, \label{cont-eqn-nond}
   \end{equation}
   \begin{equation}
   \frac{\partial u_d}{\partial t}+u_d\frac{\partial u_d}{\partial x}=-\zeta\frac{\partial \phi}{\partial x},\label{montm-eqn-nond}
   \end{equation}
   \begin{equation}
   \delta\frac{\partial f_j}{\partial t}+v\frac{\partial f_j}{\partial x}-\zeta_j\frac{m_p}{m_j}\frac{\partial \phi}{\partial x}\frac{\partial f_j}{\partial v}=0, \label{Vlasov-eqn-nond}
   \end{equation}
   \begin{equation}
   \frac{\lambda_D^2}{L^2} \frac{\partial^2\phi}{\partial x^2}=\mu_{nd}n_n-\mu_{pd}n_p-\zeta n_d, \label{Poisson-eqn-nond}
   \end{equation}
   \begin{equation}
   n_j=\sqrt{\frac{m_j}{m_p}\frac{T_p}{T_j}}\int_{-\infty}^{\infty} f_j dv, \label{density-eqn-nond}
   \end{equation}
 together with the charge neutrality condition given by 
   \begin{equation}
   \mu_{pd}+\zeta=\mu_{nd}. \label{charge-neutrality-nond}
   \end{equation}
In Eq. \eqref{Vlasov-eqn-nond}, $\zeta_j=\pm1$ for positive $(j=p)$ and negative $(j=n)$ ions and $\delta=\sqrt{z_dm_p/m_d}$. Also, in Eq. \eqref{Poisson-eqn-nond}, $\mu_{jd}=n_{j0}/z_dn_{d0}$ for $j=p$ and $n$. The basic parameters can be defined as follows: 
\begin{itemize}
\item{ $\delta\equiv\sqrt{z_dm_p/m_d}$ (for positive ions) and $m\delta$ (for negative ions with $m=m_n/m_p$), which represent the effects due to  ion inertias, and, in particular, Landau damping by both the  positive and negative ions.}
\item{$n_{d1}/n_{d0}$, the ratio of perturbed density to its equilibrium value. This measures the strength of the nonlinearity in electrostatic disturbances. } 
\item{$\lambda_D^2/L^2$: This is a measure of the strength of the wave dispersion due to deviation from the quasineutrality. Here, $L$ represents the characteristic scale length for variations of the physical quantities, namely, $n_d,~u_d,~\phi$ etc. This parameter disappears in the left-side of Eq. \eqref{Poisson-eqn-nond}, if one considers the normalization of $x$ by $\lambda_D$  instead of  $L$ . }
\end{itemize}
Note that if the ions are intertialess compared to the massive charged dusts, the ratios $\delta$ and $m\delta$ can be neglected in Eq. \eqref{Vlasov-eqn-nond}, and one can replace (hence disregarding the Landau damping effects) this equation   by the Boltzmann distributions of ions. On the other hand, when dust grains are considered cold, i.e., $T_d=0$, then the Landau damping is provided solely by the ions and the damping rate is $\propto$   $\delta$ or $m\delta$. Since one of our main interests is to study the interplay among the nonlinearity, the dispersion and the Landau damping effects, we consider \cite{ott1969} 
\begin{itemize}
\item{$\delta=\alpha_1\epsilon$,}
\item{$n_{d1}/n_{d0}=\alpha_2\epsilon$,}
\item{$\lambda_D^2/L^2=\alpha_3\epsilon$,}
\end{itemize}
 where $\epsilon(>0)$ is a smallness parameter and $\alpha_j$ ($j=1,2,3$) is a constant assumed to be of the order-unity.   Then  from Eq. \eqref{Vlasov-eqn-nond} we obtain the following   two equations for positive and negative ions as 
 \begin{equation}
 \alpha_1\epsilon\frac{\partial f_p}{\partial t}+v\frac{\partial f_p}{\partial x}-\frac{\partial\phi}{\partial x}\frac{\partial f_p}{\partial v}=0,\label{p_Vlasov-eqn-nond}
 \end{equation}
 and 
  \begin{equation}
 \alpha_1\epsilon\frac{\partial f_n}{\partial t}+v\frac{\partial f_n}{\partial x}+\frac1m\frac{\partial\phi}{\partial x}\frac{\partial f_n}{\partial v}=0.\label{n_Vlasov-eqn-nond}
 \end{equation}

 \section{Derivation of KdV equation with Landau damping}
 We note that in the limit of $\epsilon\rightarrow0$ (i.e., in the small-amplitude limit in which   ions are inertialess and the characteristic scale length  is much larger than the Debye length), Eqs. \eqref{cont-eqn-nond}-\eqref{Poisson-eqn-nond} yield the simple linear dispersion law (in nondimensional form): 
 \begin{equation}
 v_p\equiv\omega/k=\left(\mu_{pd}+\sigma\mu_{nd}\right)^{-1/2}, \label{disp-relation}
 \end{equation}
  where $\omega$ and $k$ are the wave frequency and the wave number of plane wave perturbations. Equation \eqref{disp-relation} shows that the wave becomes dispersionless with the phase speed smaller than the dust-acoustic speed $c_s$. In other words, in a frame moving at the speed $v_p$, the time derivatives of all physical quantities should vanish. Thus, for a finite $\epsilon$ with $0<\epsilon\lesssim1$, we  can expect slow variations of the wave amplitude in the moving frame of reference, and so introduce  the stretched coordinates   as \cite{taniuti1969}
 \begin{equation}
\xi=\epsilon^{1/2} (x-Mt),~\tau=\epsilon^{3/2}t, \label{stretching}
 \end{equation}
where  $M$ is the nonlinear wave speed (relative to the frame)  normalized by $c_s$, to be shown  to be equal to $v_p$ later.

  The dependent  variables are expanded in powers with respect to $\epsilon$ about the equilibrium state as   
\begin{eqnarray}
n_d&&=1+\alpha_2\epsilon n^{(1)}_d+\alpha_2^2\epsilon^2 n^{(2)}_d+\cdots,\notag \\
u_d&&=\alpha_2\epsilon u^{(1)}_d+\alpha_2^2\epsilon^2 u^{(2)}_d+\cdots,\notag \\
\phi&&=\alpha_2\epsilon\phi^{(1)}+\alpha_2^2\epsilon^2\phi^{(2)}+\cdots,\label{expantions}\\
n_j&&=1+\alpha_2\epsilon n^{(1)}_j+\alpha_2^2\epsilon^2 n^{(2)}_j+\cdots,\notag \\
f_j&&=f^{(0)}_j+\alpha_2\epsilon f^{(1)}_j+\alpha_2\epsilon^2 f^{(2)}_j+\cdots,\notag 
\end{eqnarray}
where $f^{(0)}_j$, for $j=p,n$, are assumed to be the Maxwellian given by
\begin{equation}
f^{(0)}_j=\sqrt{{1}/{2\pi}}\exp\left[\left(m_jT_p/m_pT_j\right)\left(-v^2/2\right)\right].\label{f_0 Maxwellian distribution}
\end{equation}
For convenience,   we temporarily drop the constant $\alpha_2$ in the subsequent expressions and equations. We, however, remember that $\alpha_1,~\alpha_2$ and $\alpha_3$ will explicitly appear in the coefficients of the terms associated with the Landau damping, nonlinear and the dispersion in the KdV equation. In what follows,  we substitute the expressions from Eqs. \eqref{stretching} and \eqref{expantions}  into Eqs. \eqref{cont-eqn-nond}, \eqref{montm-eqn-nond}, \eqref{Poisson-eqn-nond}, \eqref{density-eqn-nond}, \eqref{p_Vlasov-eqn-nond} and \eqref{n_Vlasov-eqn-nond} and equate successively  different powers of $\epsilon$. The results are given in the following subsections.
\subsection*{First-order perturbations and nonlinear wave speed}
\begin{figure}[ht]
\centering
\includegraphics[height=2.5in,width=3.6in]{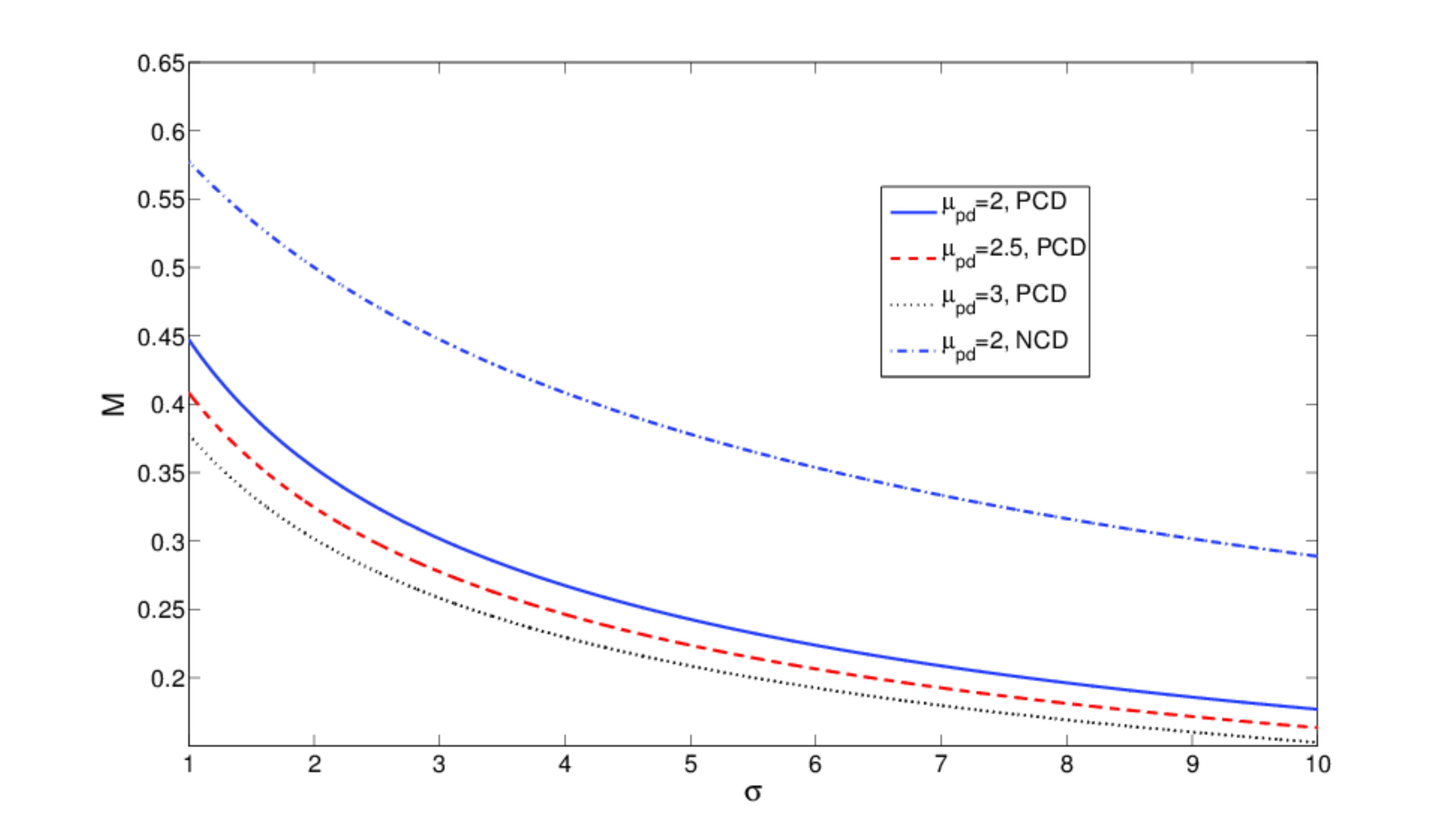}
\caption{The phase velocity $M$ of the nonlinear DAW [Eq. \eqref{phase-velocity}] is plotted against the temperature ratio $\sigma~(=T_p/T_n)$ for different values of the density ratio $\mu_{pd}~(=n_{p0}/z_dn_{d0})$ as shown in the figure. The acronym PCD (NCD) stands for the case of positively (negatively) charged dusts.}
\label{fig:fig1}
\end{figure}

Equating the coefficients of $\epsilon^{3/2}$ from Eqs. \eqref{cont-eqn-nond} and \eqref{montm-eqn-nond}, the coefficients of $\epsilon$ from Eqs. \eqref{Poisson-eqn-nond} and \eqref{density-eqn-nond},  and  the coefficients of $\epsilon^{3/2}$ from Eqs. \eqref{p_Vlasov-eqn-nond} and \eqref{n_Vlasov-eqn-nond}, we successively obtain
 \begin{equation}
 n^{(1)}_d=u^{(1)}_d/M, \label{u^(1)-n^(1)}
 \end{equation}
 \begin{equation}
 u^{(1)}_d=\zeta\frac{\phi^{(1)}}{M}, \label{u^(1)-phi^(1)}
 \end{equation}
  \begin{equation}
 0=\mu_{nd}n^{(1)}_n-\mu_{pd}n^{(1)}_p-\zeta n^{(1)}_d, \label{mu_nd-mu_bd}
 \end{equation}
 \begin{equation}
 n^{(1)}_j=\sqrt{\frac{m_j}{m_p}\frac{T_p}{T_j}}\int_{-\infty}^{\infty}f^{(1)}_jdv,\label{n^(1)_p-n^(1)_n}
 \end{equation}
  \begin{equation}
 v\frac{\partial f^{(1)}_p}{\partial\xi}+vf^{(0)}_p\frac{\partial\phi^{(1)}}{\partial\xi}=0,\label{v_p-1}
 \end{equation}
 \begin{equation}
 v\frac{\partial f^{(1)}_n}{\partial\xi}-\sigma vf^{(0)}_n\frac{\partial\phi^{(1)}}{\partial\xi}=0.\label{v_n-1}
 \end{equation}
 From Eqs.\eqref{u^(1)-n^(1)}, \eqref{u^(1)-phi^(1)} we obtain
 \begin{equation}
 n^{(1)}_d=\zeta\frac{\phi^{(1)}}{M^2}.\label{n^(1)-phi^(1)}
 \end{equation}
  Equation \eqref{v_p-1}  yields \cite{ott1969}
 \begin{equation}
 \frac{\partial f^{(1)}_p}{\partial\xi}=-f^{(0)}_p\frac{\partial\phi^{(1)}}{\partial\xi}+\lambda(\xi,\tau)\delta(v),\label{for unique soln v_p}
 \end{equation}
where $\delta(v)$ is the Dirac delta function and $\lambda(\xi, \tau)$ is an arbitrary function of $\xi$ and $\tau$. We find that the above solution for ${\partial f^{(1)}_p}/{\partial\xi}$ involves the arbitrary function $\lambda(\xi, \tau)$, and hence is not unique. Thus, for the unique solution to exist we follow Ref. \citep{ott1969} and include an extra higher-order term $\epsilon^{7/2}\alpha_1\left({\partial f^{(1)}_p}/{\partial\tau}\right)$ originating from the term $\epsilon^{5/2}\alpha_1\left({\partial f_p}/{\partial\tau}\right)$ in Eq. \eqref{p_Vlasov-eqn-nond} after the expressions  \eqref{stretching} and \eqref{expantions} being substituted. Thus, we write Eq. \eqref{v_p-1} as
\begin{equation}
\alpha_1\epsilon^2\frac{\partial f^{(1)}_{p\epsilon}}{\partial\tau}+v\frac{\partial f^{(1)}_{p\epsilon}}{\partial\xi}=-vf^{(0)}_p\frac{\partial\phi^{(1)}}{\partial\xi},\label{v_p modfd-1}
\end{equation}
Similarly, from Eq. \eqref{v_n-1}, we have 
\begin{equation}
\alpha_1\epsilon^2\frac{\partial f^{(1)}_{n\epsilon}}{\partial\tau}+v\frac{\partial f^{(1)}_{n\epsilon}}{\partial\xi}=\sigma vf^{(0)}_n\frac{\partial\phi^{(1)}}{\partial\xi}.\label{v_n modfd-1}
\end{equation}
The solutions of the initial value problems \eqref{v_p modfd-1} and \eqref{v_n modfd-1} are now unique, and can be found uniquely, once $f^{(1)}_{j\epsilon}$ for $j=p,n$ are known, by letting $\epsilon\rightarrow 0$ as
\begin{equation}
f^{(1)}_j=\lim_{\epsilon\rightarrow 0} f^{(1)}_{j\epsilon}.\label{unique sol_1}
\end{equation}
Next, taking the Fourier transform of Eq. \eqref{v_p modfd-1} with respect to $\xi$ and $\tau$ according to the formula
\begin{equation}
\hat{f}(\omega, k)=\int^{\infty}_{-\infty}\int^{\infty}_{-\infty}f(\xi, \tau)e^{i(k\xi-\omega\tau)}d\xi d\tau, \label{FT}
\end{equation}
we obtain
\begin{equation}
\hat{f}^{(1)}_{p\epsilon}=-\left(\frac{kvf^{(0)}_p}{kv-\epsilon^2\alpha_1\omega}\right)\hat{\phi}^{(1)}. \label{1FT f_p}
\end{equation}
We note that the singularity appears in Eq. \eqref{1FT f_p}. In order to avoid it   we replace $\omega$ by $\omega+i\eta$, where $\eta~(>0)$  is small, to obtain
\begin{equation}
\hat{f}^{(1)}_{p\epsilon}=-\left[\frac{kvf^{(0)}_p}{(kv-\epsilon^2\alpha_1\omega)-i\eta\alpha_1\epsilon^2}\right]\hat{\phi}^{(1)}. \label{2FT f_p}
\end{equation}
Proceeding to the limit as $\epsilon\rightarrow 0$ and using the Plemelj's formula 
\begin{equation}
\lim_{\epsilon\rightarrow 0}\frac{1}{x+i\omega}=-i\pi\delta(x)+P\left(\frac{1}{x}\right),\label{Plmj formla}
\end{equation}
where $P$ and $\delta$, respectively, denote the Cauchy principal value and the Dirac delta function, we obtain
\begin{equation}
\hat{f}^{(1)}_p=-f^{(0)}_p\hat{\phi}^{(1)},\label{3FT f_p}
\end{equation}
in which we have used the properties $xP(1/x)=1$, $x\delta(x)=0$. Next, taking Fourier inversion of Eq. \eqref{3FT f_p}, we have
\begin{equation}
f^{(1)}_p=-f^{(0)}_p\phi^{(1)}.\label{4FT f_p}
\end{equation}
Proceeding in the same way as above for the positive ions, we obtain from Eq. \eqref{v_n modfd-1} for negative ions as
\begin{equation}
f^{(1)}_n=\sigma f^{(0)}_n\phi^{(1)}.\label{4FT f_n}
\end{equation}
From Eqs. \eqref{4FT f_p} and \eqref{4FT f_n} and using Eq. \eqref{n^(1)_p-n^(1)_n} we  obtain
\begin{equation}
n^{(1)}_p=-\phi^{(1)},\label{n^(1)_p}
\end{equation}
\begin{equation}
n^{(1)}_n=\sigma\phi^{(1)}.\label{n^(1)_n}
\end{equation}
Substituting the expressions for $n_j^{(1)}$ from Eqs. \eqref{n^(1)-phi^(1)},  \eqref{n^(1)_p} and \eqref{n^(1)_n} into  Eq.\eqref{mu_nd-mu_bd}, we obtain the following expression for the nonlinear wave speed
\begin{equation}
 M=\left(\mu_{pd}+\sigma\mu_{nd}\right)^{-1/2}\equiv\left[(1+\sigma)\mu_{pd}\pm\sigma\right]^{-1/2}.\label{phase-velocity}\end{equation} 
As expected, the expression for $M$  is the same  as obtained in the linear dispersion law \eqref{disp-relation} for plane wave perturbations. The signs $\pm$ stand for the expressions according to when dust grains are considered positive or negative.  For typical laboratory \cite{kim2006,kim2013,rosenberg2007} and space plasma parameters \cite{rapp2005}, $\sigma\gtrsim1$. Also, for plasmas with positively or negatively charged dusts, the expression inside the square root is positive and/or $\mu_{pd}>1$. So, $M<1$, i.e., the nonlinear wave speed is always lower than the dust-acoustic speed. Figure \ref{fig:fig1} shows that for plasmas with positively charged dusts (See the solid, dashed and dotted lines) the value of $M$ tends to decrease with  increasing values of the positive to negative ion temperature ratio $\sigma$ as well as the density ratio $\mu_{pd}$. However, the value of $M$ becomes higher in the case of  negatively charged dusts (Compare the solid and the dash-dotted lines).
\begin{figure*}[ht]
\centering
\subfigure[]{
\includegraphics[height=2.5in,width=3.3in]{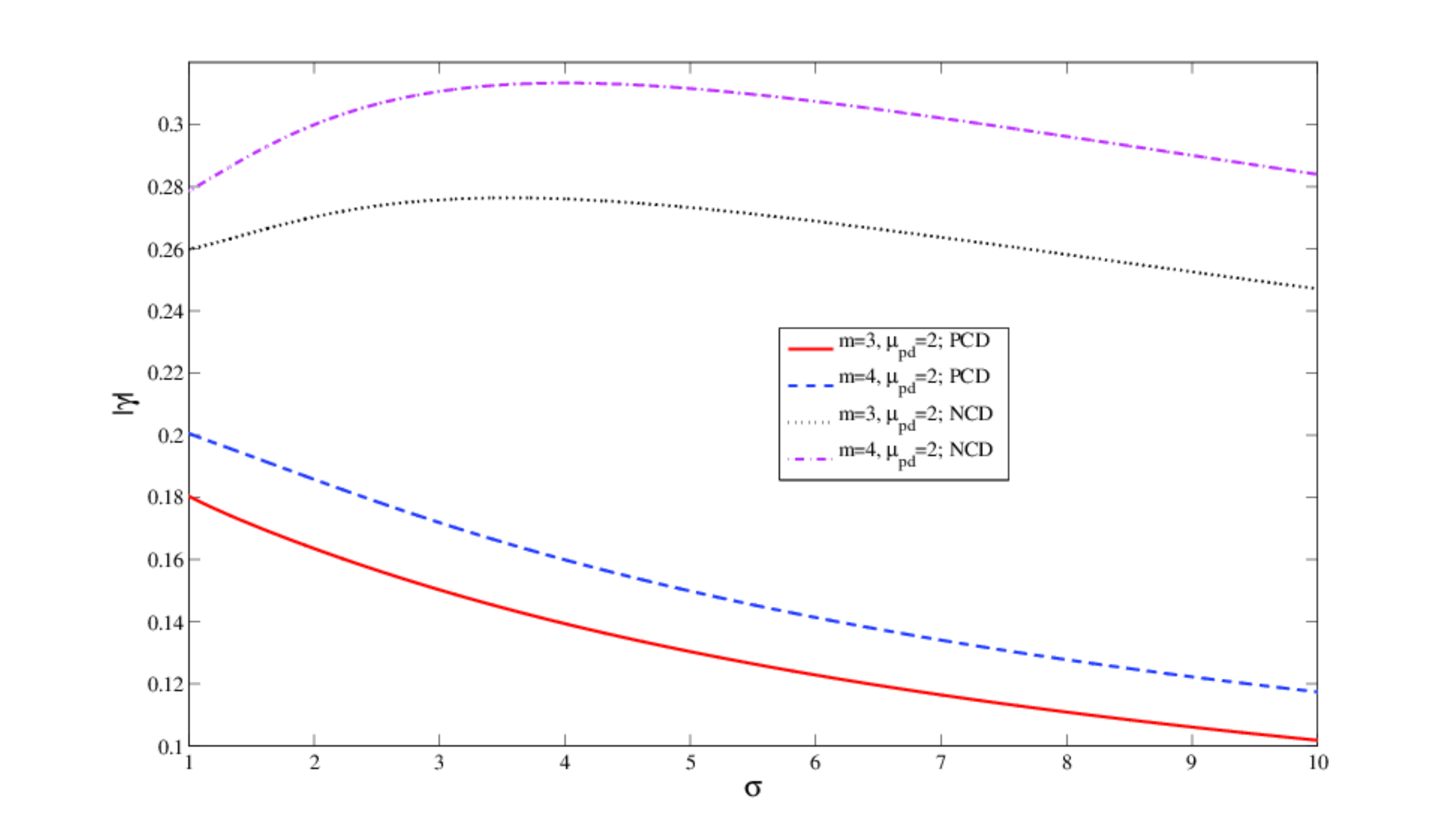}
\label{fig:fig2a}}
\quad
\subfigure[]{
\includegraphics[height=2.5in,width=3.3in]{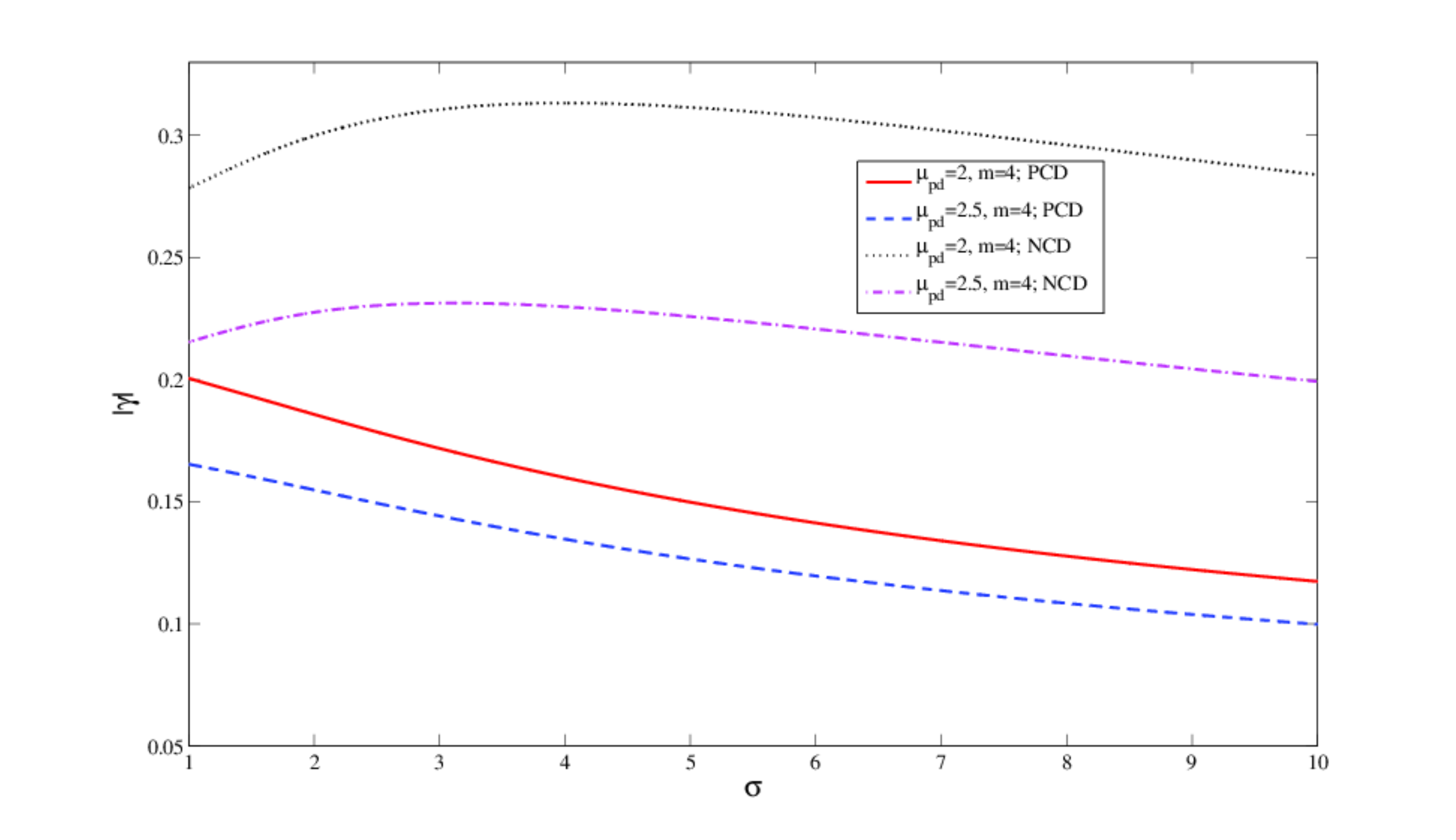}
\label{fig:fig2b}}
\caption{The linear Landau damping decrement $|\gamma|$ [Eq. \eqref{damping decrement}] is shown with the variations of the temperature ratio $\sigma~(=T_p/T_n)$ for different values of (a) the mass ratio $m~(=m_n/m_p)$ (left panel) and (b) the density ratios     $\mu_{pd}~(=n_{p0}/z_dn_{d0})$ (right panel), keeping one parameter fixed at a time, as in the figure. The acronym PCD (NCD) stands for the case of positively (negatively) charged dusts.}  
\label{fig:fig2}
\end{figure*}
\subsection*{Second-order perturbations}
 Equating the coefficients of $\epsilon^{5/2}$ from Eqs. \eqref{cont-eqn-nond} and \eqref{montm-eqn-nond}, the coefficients of $\epsilon^2$ from Eqs. \eqref{Poisson-eqn-nond} and \eqref{density-eqn-nond},  and  the coefficients of $\epsilon^{5/2}$ from Eqs. \eqref{p_Vlasov-eqn-nond} and \eqref{n_Vlasov-eqn-nond}, we successively obtain
 \begin{equation}
-M\frac{\partial n^{(2)}_d}{\partial\xi}+\frac{\partial u^{(2)}_d}{\partial\xi}+\frac{\partial n^{(1)}_d}{\partial\tau}+\frac{\partial(u^{(1)}_dn^{(1)}_d)}{\partial\xi}=0,\label{u^(2)-n^(2)}
\end{equation}
\begin{equation}
-M\frac{\partial u^{(2)}_d}{\partial\xi}+u^{(1)}_d\frac{\partial u^{(1)}_d}{\partial\xi}+\frac{\partial u^{(1)}_d}{\partial\tau}=-\zeta\frac{\partial\phi^{(2)}}{\partial\xi},\label{u^(2)-phi^(2)}
\end{equation}
\begin{equation}
\frac{\partial^2\phi^{(1)}}{\partial\xi^2}=\mu_{nd}n^{(2)}_n-\mu_{pd}n^{(2)}_p-\zeta n^{(2)}_d, \label{mu_nd-mu_bd-2}
\end{equation}
\begin{equation}
n^{(2)}_j=\sqrt{\frac{m_j}{m_p}\frac{T_p}{T_j}}\int_{-\infty}^{\infty}f^{(2)}_jdv,\label{n^(2)_p-n^(2)_n}
\end{equation}
\begin{eqnarray}
&&-\alpha_1M\frac{\partial f^{(1)}_p}{\partial\xi}+v\frac{\partial f^{(2)}_p}{\partial\xi}-\frac{\partial \phi^{(1)}}{\partial\xi}\frac{\partial f^{(1)}_p}{\partial v}\notag \\
&&+vf^{(0)}_p\frac{\partial \phi^{(2)}}{\partial\xi}=0,\label{v_p-2}
\end{eqnarray}
\begin{eqnarray}
&&-\alpha_1M\frac{\partial f^{(1)}_n}{\partial\xi}+v\frac{\partial f^{(2)}_n}{\partial\xi}+\frac1m\frac{\partial \phi^{(1)}}{\partial\xi}\frac{\partial f^{(1)}_n}{\partial v}\notag \\
&&-\sigma vf^{(0)}_n\frac{\partial \phi^{(2)}}{\partial\xi}=0.\label{v_n-2}
\end{eqnarray}
Substituting the expressions for  $f_j^{(1)}$ from Eqs.  \eqref{4FT f_p} and   \eqref{4FT f_n} into Eqs. \eqref{v_p-2} and \eqref{v_n-2} we successively obtain 
\begin{equation}
v\frac{\partial f^{(2)}_p}{\partial\xi}+vf^{(0)}_p\frac{\partial \phi^{(2)}}{\partial\xi}=(D_{pa}+vD_{pb})f^{(0)}_p,\label{v_p-3}
\end{equation}
\begin{equation}
v\frac{\partial f^{(2)}_n}{\partial\xi}-\sigma vf^{(0)}_n\frac{\partial \phi^{(2)}}{\partial\xi}=(D_{na}+vD_{nb})f^{(0)}_n,\label{v_n-3}
\end{equation} 
where
\begin{eqnarray}
&&D_{pa}=-{\alpha_1M}\frac{\partial\phi^{(1)}}{\partial\xi},~D_{pb}=\frac{1}{2}\frac{\partial(\phi^{(1)})^2}{\partial\xi}\notag\\
&&D_{na}=\alpha_1M\sigma\frac{\partial\phi^{(1)}}{\partial\xi},~D_{nb}=\frac{\sigma^2}{2}\frac{\partial(\phi^{(1)})^2}{\partial\xi}.\label{D_j(a,b)}
\end{eqnarray}
As before, to get the unique solutions for $f^{(2)}_j$ for positive $(j=p)$ and negative $(j=n)$ ions, we introduce an extra higher-order term $\epsilon^{9/2}\alpha_1\left({\partial f^{(1)}_j}/{\partial\tau}\right)$ originating from the term $\epsilon^{5/2}\alpha_1\left({\partial f_j}/{\partial\tau}\right)$ in  Eqs.\eqref{p_Vlasov-eqn-nond} and \eqref{n_Vlasov-eqn-nond} after the expressions  \eqref{stretching} and \eqref{expantions} being substituted. Thus,  we rewrite Eqs.     \eqref{v_p-3} and \eqref{v_n-3} as
\begin{eqnarray}
&&\alpha_1\epsilon^2\frac{\partial f^{(2)}_{p\epsilon}}{\partial\tau}+v\frac{\partial f^{(2)}_{p\epsilon}}{\partial\xi}+vf^{(0)}_p\frac{\partial \phi^{(2)}}{\partial\xi}\notag\\
&&=(D_{pa}+vD_{pb})f^{(0)}_p,\label{v_p-4}
\end{eqnarray}
\begin{eqnarray}
&&\alpha_1\epsilon^2\frac{\partial f^{(2)}_{n\epsilon}}{\partial\tau}+v\frac{\partial f^{(2)}_{n\epsilon}}{\partial\xi}-\sigma vf^{(0)}_n\frac{\partial \phi^{(2)}}{\partial\xi}\notag\\
&&=(D_{na}+vD_{nb})f^{(0)}_n.\label{v_n-4}
\end{eqnarray}
So, the unique solutions can be found, once $f^{(2)}_{j\epsilon}$ for $j=p,n$ are known, by letting $\epsilon\rightarrow 0$ as
\begin{equation}
f^{(2)}_j=\lim_{\epsilon\rightarrow 0} f^{(2)}_{j\epsilon}.\label{unique sol_2}
\end{equation}
Next, introducing the Fourier transform in Eq. \eqref{v_p-4} with respect to $\xi$ and $\tau$ according to the formula \eqref{FT}, we have
\begin{eqnarray}
\hat{f}^{(2)}_{p\epsilon}=-\left(\frac{kvf^{(0)}_p}{kv-\epsilon^2\alpha_1\omega}\right)\hat{\phi}^{(2)}\notag\\
-i\left(\frac{\hat{D}_{pa}+v\hat{D}_{pb}}{kv-\epsilon^2\alpha_1\omega}\right)f^{(0)}_p. \label{5FT f_p}
\end{eqnarray}
As before, to avoid the wave singularity, $\omega$ will have a small positive imaginary part. So, we replace $\omega$ by $\omega+i\eta$, where $\eta>0$, to  obtain from Eq. \eqref{5FT f_p} as
\begin{eqnarray}
\hat{f}^{(2)}_{p\epsilon}=-\left[\frac{kvf^{(0)}_p}{(kv-\epsilon^2\alpha_1\omega)-i\eta\alpha_1\epsilon^2}\right]\hat{\phi}^{(2)}\notag\\
-i\left[\frac{(\hat{D}_{pa}+v\hat{D}_{pb})}{(kv-\epsilon^2\alpha_1\omega)-i\eta\alpha_1\epsilon^2}\right]f^{(0)}_p. \label{6FT f_p}
\end{eqnarray}
Proceeding to the limit as $\epsilon\rightarrow 0$ and  using the Plemelj's formula \eqref{Plmj formla}, we have from Eq. \eqref{6FT f_p} as
\begin{eqnarray}
&&\hat{f}^{(2)}_p+f^{(0)}_p\hat{\phi}^{(2)}=-i\left[\text{P}\left(\frac{1}{kv}\right)+i\pi\frac{\text{sgn}(k)}{k}\delta(v)\right]\times\notag\\
&&(\hat{D}_{pa}+v\hat{D}_{pb})f^{(0)}_p,\label{7FT f_p}
\end{eqnarray}
where we have used the properties $x\text{P}(1/x)=1$, $x\delta(x)=0$ and $\delta(kv)=[\text{sgn}(k)/k]\delta(v)$.
We multiply both sides of Eq.\eqref{7FT f_p} by $ik$ and then integrate over $v$ to obtain
\begin{equation}
ik\left(\hat{n}^{(2)}_p+\hat{\phi}^{(2)}\right)=\hat{D}_{pb}+i\sqrt{\frac{\pi}{2}}\text{sgn}(k)\hat{D}_{pa}.\label{n^(2)_p}
\end{equation}
The  Fourier inverse transform of Eq.\eqref{n^(2)_p} yields
\begin{eqnarray}
&&\frac{\partial n^{(2)}_p}{\partial\xi}+\frac{\partial\phi^{(2)}}{\partial\xi}=\frac{1}{2}\frac{\partial(\phi^{(1)})^2}{\partial\xi}\notag\\
&&+\sqrt{\frac{\pi}{2}}F^{-1}\left[i~\text{sgn}(k)\hat{D}_{pa}\right].\label{n^(2)_p-phi^(2)}
\end{eqnarray}
Then using the convolution theorem of Fourier transform, we have from Eq. \eqref{n^(2)_p-phi^(2)}
\begin{eqnarray}
&&\frac{\partial n^{(2)}_p}{\partial\xi}+\frac{\partial\phi^{(2)}}{\partial\xi}=\frac{1}{2}\frac{\partial(\phi^{(1)})^2}{\partial\xi}\notag\\
&&+\alpha_1M\frac{1}{\sqrt{2\pi}}\text{P}\int^{\infty}_{-\infty}\frac{\partial\phi^{(1)}}{\partial\xi'}\frac{d\xi'}{\xi-\xi'},\label{n^(2)_p-phi^(2)-P}
\end{eqnarray}
where we have used $F^{-1}[i~\text{sgn}(k)]=-(1/\pi)\text{P}\left({1}/{\xi}\right)$.

Proceeding in the same way as above for positive ions,  we obtain from  Eq. \eqref{v_n-4} for negative ions as
\begin{eqnarray}
&&\frac{\partial n^{(2)}_n}{\partial\xi}-\sigma\frac{\partial\phi^{(2)}}{\partial\xi}=\frac{\sigma^2}{2}\frac{\partial(\phi^{(1)})^2}{\partial\xi}\notag\\
&&-\alpha_1Mm^{1/2}\sigma^{3/2}\frac{1}{\sqrt{2\pi}}\text{P}\int^{\infty}_{-\infty}\frac{\partial\phi^{(1)}}{\partial\xi'}\frac{d\xi'}{\xi-\xi'}.\label{n^(2)_n-phi^(2)-P}
\end{eqnarray}
\begin{figure*}[ht]
\centering
\subfigure[]{
\includegraphics[height=2.5in,width=3.3in]{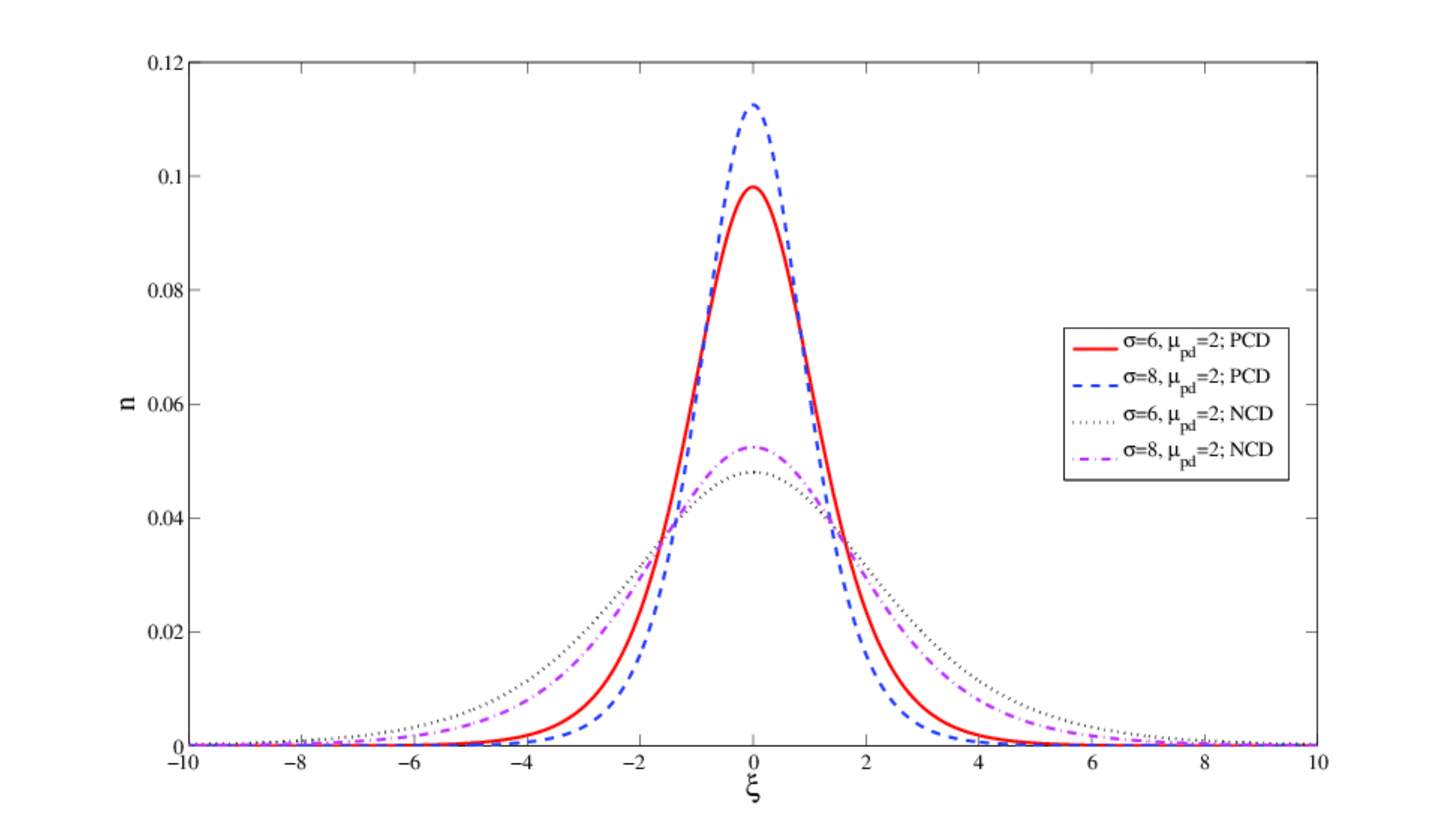}
\label{fig:fig3a}}
\subfigure[]{
\includegraphics[height=2.5in,width=3.3in]{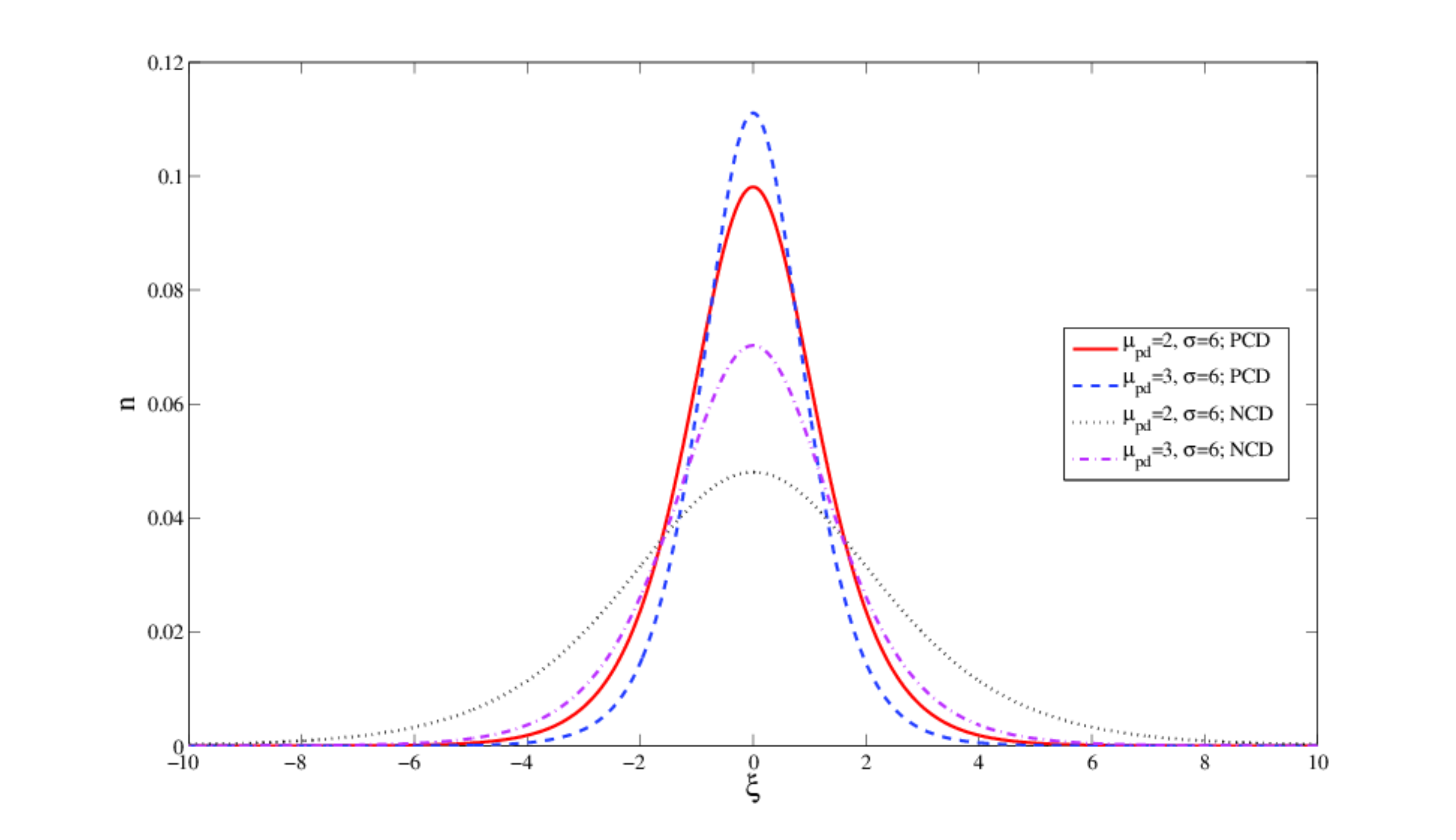}
\label{fig:fig3b}}
\caption{The KdV soliton solution $n$ [Eq. \eqref{sol of k-dv}] is shown with respect to $\xi$ at $\tau=0$  for different values of (a) the temperature  ratio  $\sigma~(=T_p/T_n)$ (left panel) and (b) the density ratio $\mu_{pd}~(=n_{p0}/z_dn_{d0})$ (right panel), keeping one parameter fixed at a time, as in the figure. The acronym PCD (NCD) stands for the case of positively (negatively) charged dusts.}
\label{fig:fig3}
\end{figure*}
\subsection*{KdV Equation with Landau damping}
In order to obtain the required KdV equation, we first eliminate ${\partial u^{(2)}_d}/{\partial\xi}$ and ${\partial n^{(2)}_d}/{\partial\xi}$ from Eqs. \eqref{u^(2)-n^(2)}-\eqref{mu_nd-mu_bd-2}   and then eliminate ${\partial n^{(2)}_j}/{\partial\xi}$  by using Eqs. \eqref{n^(2)_p-phi^(2)-P} and \eqref{n^(2)_n-phi^(2)-P}. In the resulting equation we also substitute the expressions for $n_d^{(1)}$ and $u_d^{(1)}$ from Eqs. \eqref{u^(1)-n^(1)} and  \eqref{u^(1)-phi^(1)}. Thus, we obtain the following KdV equation (Recall that the constants $\alpha_1,~\alpha_2$ and $\alpha_3$ will enter into the Landau damping, nonlinear and the dispersive terms)
\begin{equation}
\frac{\partial n}{\partial\tau}+a\text{P}\int^{\infty}_{-\infty}\frac{\partial n}{\partial\xi'}\frac{d\xi'}{\xi-\xi'} 
+bn\frac{\partial n}{\partial\xi}+c\frac{\partial^3 n}{\partial\xi^3}=0,\label{K-dV}
\end{equation}
where $n\equiv n^{(1)}_d$ and the coefficients of the   Landau damping,  nonlinear and dispersive   terms, respectively, are
\begin{equation}
a=\frac{\alpha_1}{\sqrt{8\pi}}\frac{\sigma^{-1/2}}{\sigma_1^2}\left[\zeta\sqrt{m}+(\sqrt{m}+\sigma^{-3/2})\mu_{pd}\right],\label{c}
\end{equation}
\begin{equation}
b=\frac{\alpha_2}{2\sqrt{\sigma\sigma_1}}\left[3-\zeta\frac{\sigma_2}{\sigma_1^2}\right],\label{a}
\end{equation}
\begin{equation}
c=\frac{\alpha_3}{2}(\sigma\sigma_1)^{-3/2},\label{b}
\end{equation}
with $\sigma_1=\zeta+\left(1+\sigma^{-1}\right)\mu_{pd}$,  $\sigma_2=\zeta+\left(1-\sigma^{-2}\right)\mu_{pd}$ and $\zeta=\pm1$ denoting, respectively, for positively and negatively charged dusts. 

Equation \eqref{K-dV} is the required KdV equation which describes the weakly nonlinear and weakly dispersive dust-acoustic waves in an unmagnetized dusty pair-ion  plasma with the effects of Landau damping.   Inspecting on the coefficients $a,~b$ and $c$ we find that for $\sigma>1$ and $\alpha_1,~\alpha_2,~\alpha_3\sim O(1)$, we have $\sigma_1\gtrsim\sigma_2$ and    $b>a,c$ for typical laboratory \cite{merlino1998,kim2006,kim2013}  and space \cite{rapp2005} plasma parameters.
Also,   if we set $\delta=0$, i.e. if we neglect the strength of the Landau damping associated with the   ions then   Eq.  \eqref{K-dV} reduces to the usual KdV equation which governs the small but finite amplitude nonlinear DAWs in unmagnetized dusty pair-ion plasmas.

It is of interest to examine the range of values of the parameters for which the KdV equation (without the Landau damping) is applicable to DAWs, and also to see the competition between nonlinearity and Landau damping in determining whether or not an initial wave steepens. To this end we consider typical plasma parameters that are relevant to laboratory and space plasmas. For example, for laboratory plasmas \cite{merlino1998,kim2006,kim2013} (in which the light positive ions are singly ionized potassium $K^+$ and heavy negative ions are $SF_6^-$),  we can consider $m_p=6.5\times10^{-23}$ g, $m_n=2.4\times10^{-22}$ g, $T_p=0.2~$ev$=2321~$K,
 $T_n=0.025~$ev$=290.12~$K, $n_{n0}=2\times10^9$ cm$^{-3}$,  $n_{p0}=1.3\times10^9$ cm$^{-3}$,  $n_{d0}=2\times10^6$ cm$^{-3}$, $\phi_s=0.1$ v$=3.3\times10^{-4}$ statv, 
  $R=5.04~\mu$m$=5.04\times10^{-4}~$cm and  $z_d=\sim R\phi_s/e\sim350$, so that we have [Assuming that $\alpha_1,~\alpha_2,~\alpha_3\sim O(1)$] $a=0.0412~(0.1028)$, $b=0.2719~(0.6263)$ and $c=0.0041~(0.0194)$ for positively (negatively) charged dusts. This implies that the nonlinear and the Landau damping effects are  larger than the finite Debye length (dispersive) effects so that the KdV soliton theory is  not  applicable to DAWs. This is, in fact, true for the experiment described in Ref. \citep{andersen1967}. 
  On the other hand, for space plasma parameters \cite{rapp2005} (e.g., a dusty region at an altitude of about 95 km) in which $m_p=28m_{\text{proton}}=4.7\times10^{-23}$ g, $m_n=300m_{\text{proton}}=5.02\times10^{-22}$ g, $T_p=200~$K, $T_n=200~$K, $n_{n0}=2\times10^6$ cm$^{-3}$,  $n_{p0}=10^6$ cm$^{-3}$, $z_dn_{d0}=10^6$ cm$^{-3}$, $\phi_s=0.7$ v$=2.3\times10^{-3}$ statv,  $R=0.6~$nm$=6\times10^{-8}~$cm, we have [Assuming that $\alpha_1,~\alpha_2,~\alpha_3\sim O(1)$] $a=0.1670~(0.1995)$, $b=0.8340~(1)$  and $c=0.09~(0.5)$ for positively (negatively) charged dusts. In this case, though the Landau damping coefficient is lower than the nonlinear one, but is larger than or comparable with the dispersive coefficient. Thus, both in laboratory and space plasma environments the Landau damping effects on DAWs can no longer be negligible, but may play crucial roles in reducing the wave amplitude which will be shown shortly.

 Next, to obtain the regular Landau damping of DAWs in plasmas we set $b=c=0$. Then Eq. \eqref{K-dV} reduces to
\begin{equation}
\frac{\partial n}{\partial\tau}+a\text{P}\int^{\infty}_{-\infty}\frac{\partial n}{\partial\xi'}\frac{d\xi'}{\xi-\xi'}=0.\label{regular LD}
\end{equation}
Taking the Fourier transform of Eq. \eqref{regular LD} according to the formula \eqref{FT} and using the result that the inverse transform of $[i~\text{sgn}(k)]$ is $-(1/\pi)\text{P}\left(1/\xi\right)$, we have 
\begin{eqnarray}
\omega=&&-ik\alpha_1\sqrt{\frac{\pi}{8}}\frac{\sigma^{-1/2}}{\sigma_1^2}\left[\zeta\sqrt{m}+(\sqrt{m}+\sigma^{-3/2})\mu_{pd}\right]\notag\\
\equiv&&-i\pi ka.\label{Dispersion relation}
\end{eqnarray}
Thus, the DAWs become  damped due to the finite positive and negative ion inertial effects as $\alpha_1\propto\sqrt{m_j/m_d}$, and the  damping decrement (nondimensional) $\gamma$   is given by 
\begin{eqnarray}
|\gamma|\sim&&\sqrt{\frac{m_pz_d}{m_d}}\sqrt{\frac{\pi}{8}}\frac{\sigma^{-1/2}}{\sigma_1^2}\left[\zeta\sqrt{m}+(\sqrt{m}+\sigma^{-3/2})\mu_{pd}\right]\notag\\
\equiv&&\pi a.\label{damping decrement}
\end{eqnarray}
The variations of $|\gamma|$ with respect to $\sigma$ is shown in Fig. \ref{fig:fig2} for different values of the mass ratio  $m$ [Fig. \ref{fig:fig2a}] and the density ratio $\mu_{pd}$ [Fig. \ref{fig:fig2b}]. It is seen that  the value of  $|\gamma|$ slowly decreases with an increase of the temperature ratio $\sigma$ for plasmas with positively charged dusts. However, for plasmas with negatively charged dusts, the value of $|\gamma|$   initially increases until $\sigma$ reaches its critical value, and then decreases with increasing values of $\sigma$.  These may be   consequences of typical laboratory plasmas (see above) in which positive ion temperature is higher than the neagtive ions.  Also, as the ratio $m$ $(\mu_{pd})$ increases, the value of $|\gamma|$ increases (decreases) [See the solid and dashed lines for positively charged dusts, and the dotted and dash-dotted lines for negatively charged dusts]. Thus, dusty plasmas with a higher concentration of  positive ions (and hence that of the negative  ions   in order to maintain the charge neutrality) than the charged dusts reduce the linear Landau damping rate of DAWs. This may be true for some laboratory plasmas as described above. Also, since the mass difference of positive and negative ions can be higher in space plasmas as mentioned above, an enhancement of the damping rate is more likely to occur there.   Furthermore,   a higher value  of $|\gamma|$  for plasmas with negatively charged dusts  is seen to occur (Compare the solid and dotted lines or the dotted and dash-dotted lines).

\begin{figure*}[ht]
\centering
\subfigure[]{
\includegraphics[height=2.2in,width=3.2in]{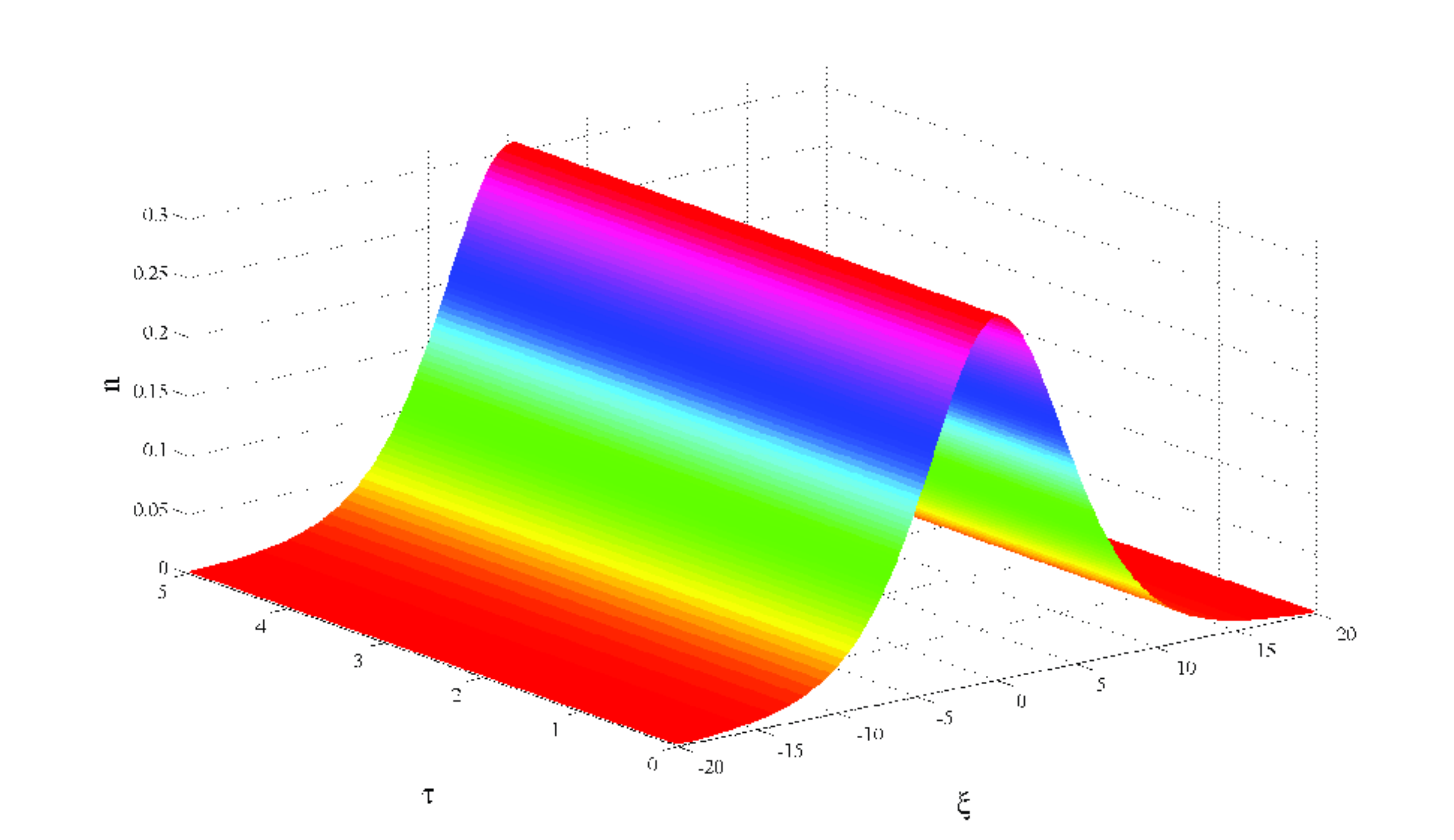}
\label{fig:fig4a}}
\quad
\subfigure[]{\includegraphics[height=2.2in,width=3.2in]{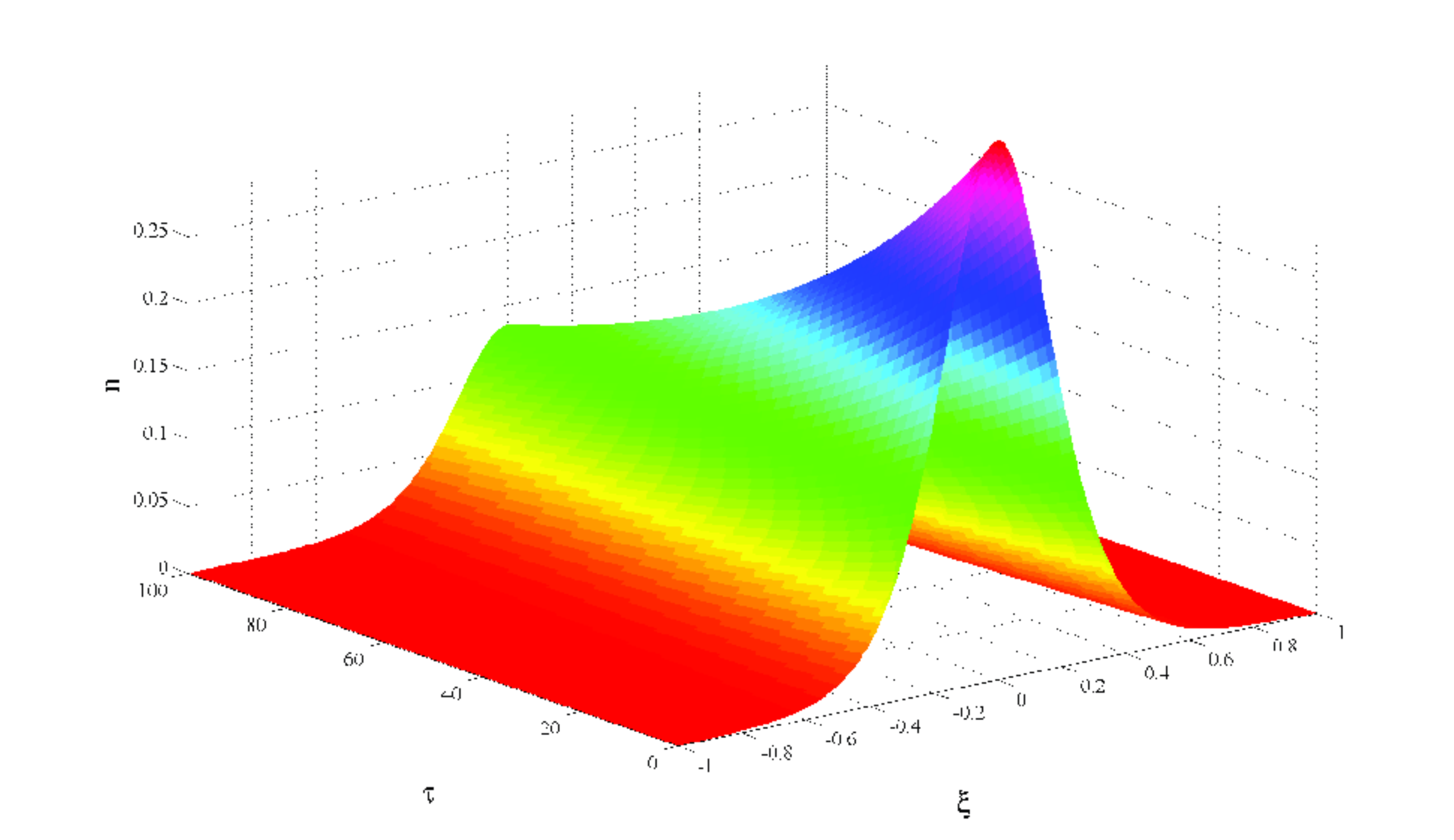}
\label{fig:fig4b}}
\caption{Typical forms of the soliton solutions given by Eqs. \eqref{sol of k-dv} and \eqref{final-sol} are shown. The left (right) panel shows the solution without (with) the Landau damping effect. The parameter values are for plasmas with positively charged dusts with $m=2,~\mu_{pd}=1.2$,  $\sigma=1.2$, $\alpha_1=0.6,~\alpha_2=0.8$ and $\alpha_3=8$ so that $b\sim c\gg a$. Also, for the left panel (a) $U_0=0.06$, and for the right panel (b) $U_0=40$ and $N_0=0.4$.  }
\label{fig:fig4}
\end{figure*}
\section{Solitary wave solution of the KdV equation with Landau damping}
We  note that in absence of  the Landau damping effect (i.e. $a=0$),  Eq. \eqref{K-dV} reduces to the usual KdV equation, the solitary wave solution of which   is given by
\begin{equation}
n=N~\text{sech}^2\left(\frac{\xi-U_0\tau}{W}\right),\label{sol of k-dv}
\end{equation}
where $N=3U_0/b$ is the amplitude and $W=\left({12c}/{Nb}\right)^{1/2}\equiv\sqrt{4c/U_0}$ is the width and $U_0=Nb/3$ is the constant phase speed (normalized by $c_s$) of the solitary wave. 

To find the solitary wave solution of Eq. \eqref{K-dV} with the effect of a small amount of Landau damping, we follow Ref. \cite{ott1969}. Thus, integrating Eq. \eqref{K-dV} with respect to $\xi$  one can obtain 
\begin{equation}
\frac{\partial}{\partial\tau}\int^{+\infty}_{-\infty}n~d\xi=0,
\end{equation}
i.e., Eq. \eqref{K-dV} conserves the total number of particles.  Furthermore, multiplying Eq. \eqref{K-dV} by $n$ and integrating over $\xi$ yields
\begin{equation}
\frac{\partial}{\partial\tau}\int_{-\infty}^{\infty}n^2(\xi, \tau) d\xi\leq 0,\label{positive definite, H theorem}
\end{equation}
where the equality sign holds only when $n=0$ for all $\xi$. Equation \eqref{positive definite, H theorem}   states that an initial perturbation of the form \eqref{sol of k-dv}  for which
\begin{equation}
\int^{+\infty}_{-\infty}n^2~d\xi<\infty,
\end{equation}
 will decay to zero. That is, the wave amplitude $N$ is not a constant but decreases slowly with time. In what follows we perform a perturbation analysis of Eq. \eqref{K-dV} assuming that $a~(\gg\epsilon)$ is a small parameter with $1\sim b\sim c\gg a$. The latter may be satisfied for plasmas (e.g., laboratory plasmas in which pair-ions can be $Ar^+SF_6^-$) with $m>1$ and $\mu_{pd}\sim\sigma\sim1$.  So,    we introduce a new space coordinate $z$ in a frame moving with the solitary wave and normalized to its width as 
\begin{equation}
z=\left(\xi-\frac {b}3\int_{0}^{\tau}Nd\tau\right)/W,\label{space coordinate}
\end{equation}
where $N$ is assumed to vary slowly with time and $N=N(a, \tau)$. Also, assume that $n\equiv n(z, \tau)$. Under this transformation Eq. \eqref{K-dV} becomes
\begin{eqnarray}
&&\frac{\partial n}{\partial\tau}+\frac{a}{W}P\int_{-\infty}^{\infty}\frac{\partial n}{\partial z'}\frac{dz'}{z-z'}-\left[\frac{Nb}{3W}-\frac{z}{2N}\left(\frac{dN}{d\tau}\right)\right]\frac{\partial n}{\partial z}\notag\\
&&+\frac{b}{W}n\frac{\partial n}{\partial z}+\frac{c}{W^3}\frac{\partial^3n}{\partial z^3}=0,\label{after subst SP}
\end{eqnarray}
where we have used ${\partial n}/{\partial z'}={\partial n}/{\partial z}$ at $z=z'$. 

 Next, to investigate the solution of Eq. \eqref{after subst SP}, we follow Ref. \citep{ott1969} and generalize the multiple time scale analysis with respect to $a$. Thus,  we consider the solution as \cite{bandyo2002a,bandyo2002b}
\begin{equation}
n(z, \tau)=n^{(0)}+an^{(1)}+a^2n^{(2)}+a^3n^{(3)}+\cdots,\label{multiple time scale}
\end{equation}
where $n^{(i)},~i=0, 1, 2, 3,\cdots$, are functions of $\tau=\tau_0, \tau_1, \tau_2, \tau_3,\cdots$ in which $\tau_i$ are given by
\begin{equation}
\tau_i=a^{i}\tau,\label{tau_i}
\end{equation}
where $i=0, 1, 2, 3,\cdots$.
Substituting \eqref{multiple time scale} into Eq. \eqref{after subst SP}, we obtain
\begin{eqnarray}
&&\left(\frac{\partial n^{(0)}}{\partial\tau}+a\frac{\partial n^{(0)}}{\partial\tau_1}+\cdots\right)+a\left(\frac{\partial n^{(1)}}{\partial\tau}+a\frac{\partial n^{(1)}}{\partial\tau_1}+\cdots\right)\notag\\
&&+\left[-\frac{Nb}{3W}+\frac{z}{2N}\left(\frac{\partial N}{\partial\tau}+a\frac{\partial N}{\partial\tau_1}+\cdots\right)\right]\times\notag\\
&&\frac{\partial}{\partial z}\left(n^{(0)}+an^{(1)}+\cdots\right)+\frac{a}{W}P\int_{-\infty}^{\infty}\frac{\partial n^{(0)}}{\partial z'}\frac{dz'}{z-z'}\notag\\
&&+\frac{b}{W}n^{(0)}\frac{\partial n^{(0)}}{\partial z}+\frac{ab}{W}\left(n^{(1)}\frac{\partial n^{(0)}}{\partial z}+n^{(0)}\frac{\partial n^{(1)}}{\partial z}\right)\notag\\
&&+\frac{c}{W^3}\frac{\partial^3n^{(0)}}{\partial z^3}+\frac{ac}{W^3}\frac{\partial^3n^{(1)}}{\partial z^3}+\cdots=0,\label{big equation}
\end{eqnarray}
Equating the coefficients of zeroth and first-order of $a$, we successively obtain  from Eq. \eqref{big equation} as
\begin{equation}
\beta\left[\frac{\partial}{\partial\tau}+\frac{z}{2N}\frac{\partial N}{\partial\tau}\frac{\partial}{\partial z}\right]n^{(0)}+M\frac{\partial n^{(0)}}{\partial z}=0,\label{0th order eq}
\end{equation}
\begin{equation}
\beta\left[\frac{\partial}{\partial\tau}+\frac{z}{2N}\frac{\partial N}{\partial\tau}\frac{\partial}{\partial z}\right]n^{(1)}+\frac{\partial}{\partial z} Mn^{(1)}=\beta Rn^{(0)},\label{1st order eq}
\end{equation}
where
\begin{equation}
\beta=\frac{W^3}{c}=24\sqrt{\frac{3c}{{b}^3}}N^{-3/2},\label{beta}
\end{equation}
\begin{equation}
M=\frac{\partial^2}{\partial z^2}+4\left(3\frac{n^{(0)}}{N}-1\right),\label{M}
\end{equation}
\begin{eqnarray}
Rn^{(0)}=-\left[\frac{\partial n^{(0)}}{\partial\tau_1}+\frac{z}{2N}\frac{\partial N}{\partial \tau_1}\frac{\partial n^{(0)}}{\partial z}\right.\notag\\
\left.+\frac{1}{W}\text{P}\int_{-\infty}^{\infty}\frac{\partial n^{(0)}}{\partial z'}\frac{dz'}{z-z'}\right].\label{Rq^(0)}
\end{eqnarray}\\
Next, imposing the boundary conditions, namely, $n^{(0)}$, ${\partial n^{(0)}}/{\partial z}$, ${\partial^2 n^{(0)}}/{\partial z^2}\rightarrow 0$ as $z\rightarrow\pm\infty$, it can easily be shown that $n^{(0)}=N~\text{sech}^2z$ is the soliton solution of the equation $M{\partial n^{(0)}}/{\partial z}=0$. Hence $n^{(0)}=N~\text{sech}^2z$ will be the soliton solution of Eq. \eqref{0th order eq} if and only if \cite{bandyo2002a,bandyo2002b}
\begin{equation}
\frac{\partial N}{\partial\tau}=0.\label{condition 1}
\end{equation}
Under the condition \eqref{condition 1}, Eq. \eqref{1st order eq} reduces to
\begin{equation}
\beta\left[\frac{\partial n^{(1)}}{\partial\tau}\right]+\frac{\partial}{\partial z} \left(Mn^{(1)}\right)=\beta Rn^{(0)}.\label{1st order eq-AC}
\end{equation}\\
In order that the solution of Eq. \eqref{1st order eq-AC} exists, it is necessary that $Rn^{(0)}$ be orthogonal to all solutions $g(z)$, of $L^{+}[g]=0$ which satisfy $g(\pm\infty)=0$, where $L^{+}$ is the operator adjoint to $L$, defined by 
\begin{equation}\\
\int_{-\infty}^{\infty}\psi_1(z)L[\psi_2(z)]dz=\int_{-\infty}^{\infty}\psi_2(z)L^{+}[\psi_1(z)]dz,\label{adjt oprtr}
\end{equation}
where $\psi_1(\pm\infty)=\psi_2(\pm\infty)=0$, and  the only solution of $L^{+}[g]=0$ is $g(z)=\text{sech}^2z$. Thus, we have
\begin{equation}
\int_{-\infty}^{\infty} Rn^{(0)}\text{sech}^2z dz=0,\label{eqn}
\end{equation}
which gives
\begin{eqnarray}
&&\frac{\partial N}{\partial \tau}+\frac12a\sqrt{\frac{b}{3c}}N^{3/2}\times\notag\\
&&\text{P}\int_{-\infty}^{\infty}\int_{-\infty}^{\infty}\frac{\text{sech}^2z}{z-z'}\frac{\partial}{\partial z'}\left(\text{sech}^2z'\right)dzdz'=0.\label{1st ordr diffnal eq}
\end{eqnarray}
Equation \eqref{1st ordr diffnal eq}  is a first-order differential equation for the solitary wave amplitude $N(a, \tau)$, the solution of which is 
\begin{equation}
N(a, \tau)=N_0\left(1+\frac{\tau}{\tau_0}\right)^{-2},\label{solution}
\end{equation}
where $N=N_0$   at $\tau=0$ and $\tau_0$ is given by
\begin{equation}
\tau_0^{-1}=\frac{a}{4}\sqrt{\frac{bN_0}{3c}}\text{P}\int_{-\infty}^{\infty}\int_{-\infty}^{\infty}\frac{\text{sech}^2z}{z-z'}\frac{\partial}{\partial z'}\left(\text{sech}^2z'\right)dzdz'.\label{tau'}
\end{equation}
Taking Fourier transform of $\text{sech}^2z$ and making use of the identity 
\begin{equation}
\text{P}\int_{-\infty}^{\infty}\frac{\exp(ikz)}{z-z'}dz=i\pi~\text{sgn}~\kappa~\exp(i\kappa z')
\end{equation}
one obtains
\begin{equation}
\text{P}\int_{-\infty}^{\infty}\int_{-\infty}^{\infty}\frac{\text{sech}^2z}{z-z'}\frac{\partial}{\partial z'}\left(\text{sech}^2z'\right)dzdz'=\frac{24}{\pi^2}\zeta(3)=2.92, 
\end{equation}
 where $\zeta$ is the Riemann zeta function. Thus,   from Eq. \eqref{tau'} we have
\begin{equation}
\tau_0\approx\frac{1.37}{a}\sqrt{\frac{3c}{bN_0}}.\label{tau-final}
\end{equation}
The final soliton solution of Eq. \eqref{K-dV} with the Landau damping is  then given by
\begin{eqnarray}
n=&&N_0\left(1+\frac{\tau}{\tau_0}\right)^{-2}~\text{sech}^2\left[\left(\xi-\frac {b}3\int_{0}^{\tau}Nd\tau\right)/W\right]\notag\\
&&+O(a).\label{final-sol}
\end{eqnarray}
This shows that the amplitude of DA solitary waves decays slowly with time with the effect of a   small amount of the Landau damping. From Eq. \eqref{final-sol}, it is also seen that as the wave amplitude decreases, the propagation speed also slows down in widening the pulse width. 
\begin{figure}[ht]
\centering
 \includegraphics[height=2.5in,width=3.6in]{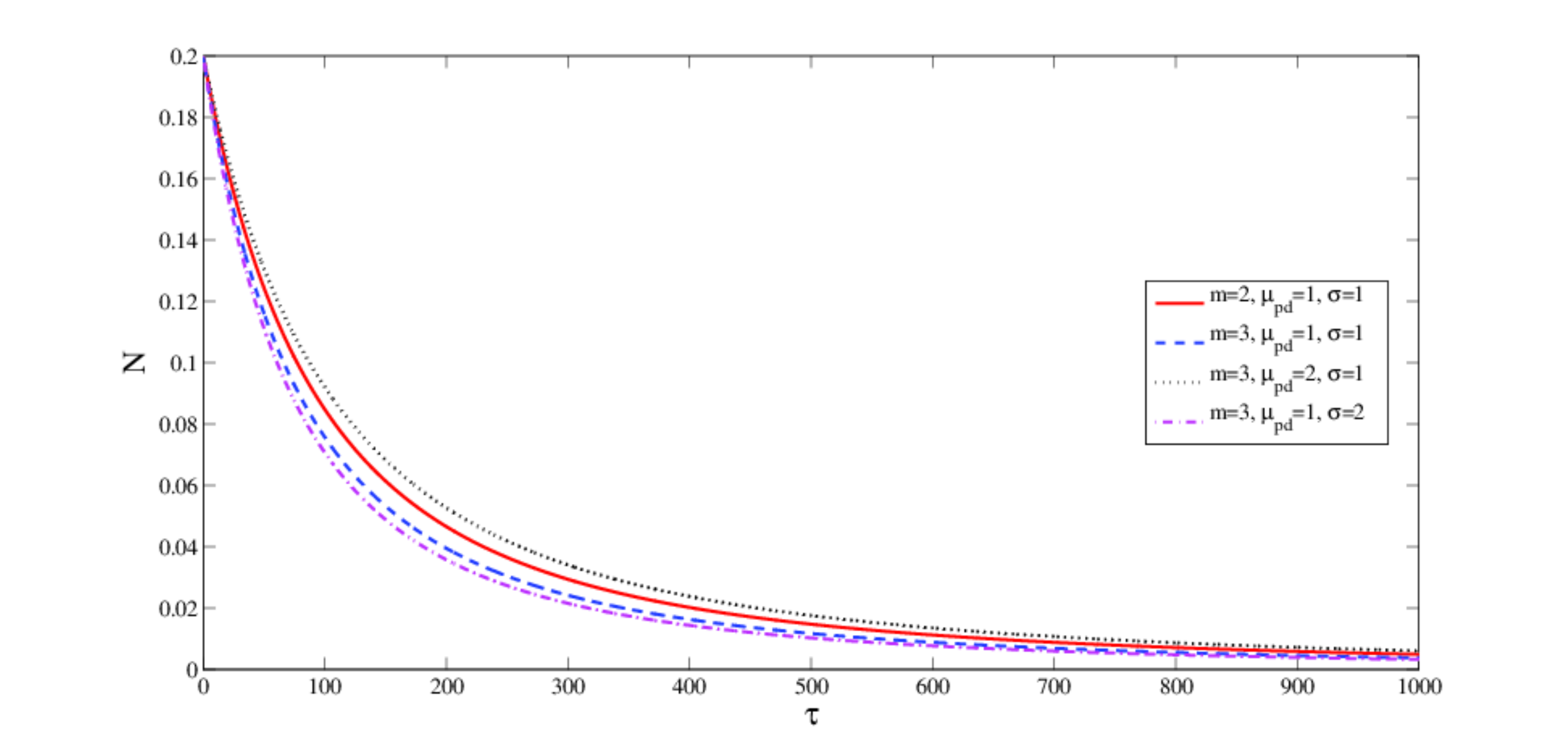}
\caption{The decay of the solitary wave amplitude [Eq. \eqref{solution}] with time is shown as a result of the Landau damping effect for different values of the   parameters $m,~\mu_{pd}$ and $\sigma$  as relevant  for plasmas with positively charged dusts. The other values are taken as $\alpha_1=0.6,~\alpha_2=0.8$, $\alpha_3=8$ and $N_0=0.2$ so that $b\sim c\gg a$ is satisfied.  The qualitative behaviors remain the same for plasmas with  negatively charged dusts.  }
\label{fig:fig5} 
\end{figure}

We numerically investigate the properties of the soliton solutions given by Eqs. \eqref{sol of k-dv} and \eqref{final-sol} with different plasma parameters. Figure \ref{fig:fig3} exhibits the characteristics  of the KdV soliton for different values of (a) $\sigma$ [Fig. \ref{fig:fig3a}] and (b) $\mu_{pd}$ [Fig. \ref{fig:fig3b}]   for plasmas with positively (PCD) and negatively charged dusts (NCD). The changes of values of the amplitude and width are more pronounced in case of plasmas with negatively charged dusts. In this case, the solitons  get widened and  their   amplitudes   remain smaller than the positively charged dust case. From Fig. \ref{fig:fig3a}  it is seen that as the temperature ratio $\sigma$ increases, both the amplitude and width of the soliton decrease. On the other hand, as the ratio $\mu_{pd}$ increases,  an increase of the amplitude and a decrease of the width are found to occur [Fig. \ref{fig:fig3b}].     

Typical forms of the KdV soliton [Fig. \ref{fig:fig4a}] and soliton with the effect of Landau damping [Fig. \ref{fig:fig4b}] are shown in Fig. \ref{fig:fig4}. The parameter values are for typical laboratory plasmas (as mentioned before) with positively charged dusts. The features are also similar in the case of negatively charged dusts. Clearly, the wave amplitude slows down with time by the effect of a small amount of Landau damping. Such a decay of the wave amplitude  with time [Eq. \eqref{solution}] is also exhibited  in Fig. \ref{fig:fig5} for different parameter values. The latter are   for  plasmas with positively charged dusts with $m>1$ and $\mu_{pd}\sim\sigma\sim1$. We find that as the mass and temperature ratios increase (Relevant for typical laboratory dusty plasmas mentioned above) there is a slowing down of the wave amplitude. However, such a reduction of the amplitude  is more pronounced with a small enhancement of $\sigma$. Furthermore, an increase of the wave amplitude with the density ratio $\mu_{pd}$ is also found to occur. 

\section{Conclusion}
We have investigated the Landau damping effects of both positive and negative ions on small but finite amplitude electrostatic solitary waves in dusty negative-ion plasmas consisting of mobile charged dusts and both positive and negative ions. A Korteweg de-Vries (KdV) equation with a nonlocal integral term (Landau damping) is derived   which governs the dynamics of  weakly nonlinear and weakly dispersive DAWs.   It is found that for typical laboratory \cite{merlino1998,kim2006,kim2013} and space plasmas \cite{rapp2005}, the Landau damping (and the nonlinear) effects for both positive and negative ions become dominant over the finite Debye length (dispersive) effects for which the KdV soliton theory is no longer applicable to DAWs in dusty pair-ion plasmas.   In such cases and in presence of Landau damping, the soliton amplitude is found to decay with time. The wave amplitude also decreases with an increase of the ratios of negative to positive ion masses $(m)$ and the  positive to negative ion temperatures $(\sigma)$. However, the amplitude may be  increased with increasing values of the positive ion to dust number density ratio $(\mu_{pd})$. On the other hand,  the amplitude and width of the KdV soliton  are found to decrease with an increase of $\sigma$, whereas its amplitude increases and the width decreases with an increase of $\mu_{pd}$.  The results should be useful for understanding the evolution of dust-acoustic solitary waves in dusty negative ion plasmas such as those in laboratory \cite{merlino1998,kim2006,kim2013} and space environments \cite{geortz1989,rapp2005}.
\section*{Acknowledgement}
{A. B. is thankful to University Grants Commission (UGC), Govt. of India, for Rajib Gandhi National Fellowship with Ref. No. F1-17.1/2012-13/RGNF-2012-13-SC-WES-17295/(SA-III/Website).  This research was partially supported by   the SAP-DRS (Phase-II), UGC, New Delhi, through sanction letter No. F.510/4/DRS/2009 (SAP-I) dated 13 Oct., 2009, and by the Visva-Bharati University, Santiniketan-731 235, through Memo No.  REG/Notice/156 dated January 7, 2014.}


\begin{thebibliography}{50}
\bibitem{shukla2002}  P. K. Shukla and A. A. Mamun, {\it Introduction to Dusty plasma Physics} (Institute of Physics Publishing, Bristol, 2002).
\bibitem{geortz1989} G. K. Geortz, Rev. Geophys. {\bf27}, 271 (1989).
\bibitem{merlino1998} R. L. Merlino, A. Barkan, C. Thompson, and N. D'. Angelo, Phys. Plasmas {\bf5}, 1607 (1998).
\bibitem{narcisi1971} R. S. Narcisi, A. D. Bailey, L. D. Lucca, C. Sherman, and D. M. Thomas, J. Atmos. Terr. Phys. {\bf 33}, 1147 (1971).
\bibitem{rapp2005}   M. Rapp, J. Hedin, and I. Strelnikova, M. Friedrich, J. Gumbel, 
and F.-J. L{\"u}bken, Geophys. Res. Lett. \textbf{32}, L23821 (2005). 
\bibitem{choi1993} S. J. Choi and M. J. Kushner, J. Appl. Phys. {\bf74}, 853 (1993).
\bibitem{yabe1994} E. Yabe and K. Takahashi, Appl. Phys. Lett. {\bf 65}, 694 (1994).
\bibitem{kim2006} S. H. Kim and R. L. Merlino, Phys. Plasmas \textbf{13}, 052118 (2006). 
\bibitem{kim2013} S. H. Kim,  R. L. Merlino, J. K. Meyer, and M. Rosenberg,  J. Plasma Phys.  \textbf{79}, 1107 (2013).
\bibitem{rosenberg2007} M. Rosenberg and R. L. Merlino, Planet. Space Sci. \textbf{55}, 1464 (2007).
\bibitem{misra2012} A. P. Misra, N. C. Adhikary, and P. K. Shukla, Phys. Rev. E \textbf{86},   056406 (2012).
\bibitem{misra2013} A. P. Misra and N. C. Adhikary, Phys. Plasmas \textbf{20}, 102309 (2013).
\bibitem{rehman2012} H. U. Rehman, Chin. Phys. Lett. \textbf{29}, 65201 (2012).
\bibitem{ghosh2013} S. Ghosh, N. Chakrabarti, M. Khan, and M. R. Gupta, PRAMANA-J. Phys. \textbf{80}, 283 (2013).
\bibitem{oohara2005} W. Oohara, D. Date, and R. Hatakeyama, Phys. Rev. Lett. \textbf{95}, 175003 (2005).
\bibitem{saleem2007} H. Saleem, Phys. Plasmas \textbf{14}, 014505 (2007).
\bibitem{reid1990} G. C. Reid, J. Geophys. Res. {\bf95}, 13891 (1990).
\bibitem{rao1990} N. N. Rao, P. K. Shukla, M. Y. Yu, Planet Space Sci. \textbf{38}, 543 (1990).
\bibitem{landau1946} L. Landau, J. Phys. (USSR) \textbf{10}, 25 (1946).
\bibitem{ott1969} E. Ott and R. N. Sudan, Phys. Fluids {\bf12}, 2388 (1969).
\bibitem{ott1970} E. Ott and R. N. Sudan, Phys. Fluids {\bf13}, 1432 (1970).
\bibitem{bandyo2002a} A. Bandyopadyay and K. P. Das, Phys. Plasmas \textbf{9}, 465 (2002).
\bibitem{bandyo2002b} A. Bandyopadyay and K. P. Das, Phys. Plasmas \textbf{9}, 3333 (2002).
\bibitem{ghosh2011} S. Ghosh and R. Bharuthram, Astrophys. Space Sci. {\bf331}, 163 (2011).
\bibitem{taniuti1969} T. Taniuti and N. Yajima, J. Math. Phys. {\bf10}, 1369 (1969).
\bibitem{andersen1967} H. K. Andersen, N. D'Angelo, P. Michelsen, and P. Nielsen, Phys. Fluids {\bf11}, 149 (1967).



\end{thebibliography}
\end{document}